\documentclass[superscriptaddress,aps,pra,twocolumn]{revtex4}
\usepackage{bm}
\usepackage{ulem}
\usepackage{epsfig}
\usepackage{graphicx}
\usepackage{amssymb,amsmath,amsbsy,amsgen,amsfonts}    
\usepackage{dcolumn}
\usepackage{amsthm}
\usepackage{mathrsfs}
\usepackage{latexsym}
\usepackage{array}
\usepackage{color}
\usepackage{amstext}
\allowdisplaybreaks[1]
\usepackage{txfonts}

\usepackage{epstopdf} 

\newcommand{\bra}[1]{\langle{#1}\vert}
\newcommand{\ket}[1]{\vert{#1}\rangle}

\newcommand{\be}{\begin{equation}}
\newcommand{\ee}{\end{equation}}
\newcommand{\ba}{\begin{array}}
\newcommand{\ea}{\end{array}}
\newcommand{\bqa}{\begin{eqnarray}}
\newcommand{\eqa}{\end{eqnarray}}

\DeclareSymbolFont{symbols}{OMS}{cmsy}{m}{n}

\begin{document}

\title{Probing the effect of interaction on Anderson localization using linear photonic lattices }
\author{Changhyoup Lee}
\email{changdolli@gmail.com}
\affiliation{Centre for Quantum Technologies, National University of Singapore, 3 Science Drive 2, Singapore 117543}
\author{Amit Rai}
\affiliation{Centre for Quantum Technologies, National University of Singapore, 3 Science Drive 2, Singapore 117543}
\author{Changsuk Noh}
\affiliation{Centre for Quantum Technologies, National University of Singapore, 3 Science Drive 2, Singapore 117543}
\author{Dimitris G. Angelakis}
\email{dimitris.angelakis@gmail.com}
\affiliation{Centre for Quantum Technologies, National University of Singapore, 3 Science Drive 2, Singapore 117543}
\affiliation{School of Electronic and Computer Engineering, Technical University of Crete, Chania, Greece 73100}

\date{\today}

\begin{abstract}
We show how two-dimensional waveguide arrays can be used to probe the effect of on-site interaction on Anderson localization of two interacting bosons in one dimension. 
It is shown that classical light and linear elements are sufficient to experimentally probe the interplay between interaction and disorder in this setting. For experimental relevance, we evaluate the participation ratio and the intensity correlation function as measures of localization for two types of disorder (diagonal and off-diagonal), for two types of interaction (repulsive and attractive), and for a variety of initial input states. Employing a commonly used set of initial states, we show that the effect of interaction on Anderson localization is strongly dependent on the type of disorder and initial conditions, but is independent of whether the interaction is repulsive or attractive. We then analyze a certain type of entangled input state where the type of interaction is relevant and discuss how it can be naturally implemented in waveguide arrays. We conclude by laying out the details of the two-dimensional photonic lattice implementation including the required parameter regime.
\end{abstract}


\maketitle

{\it Introduction.} 
Anderson localization (AL)~\cite{Anderson58}, one of the most famous manifestations of quantum destructive interference, has been probed and verified in perhaps the most diverse physical platforms such as light propagation in spatially random optical media \cite{Schwartz07, Segev13}, noninteracting Bose-Einstein condensates in random optical potentials \cite{Billy08, Roati08}, microwave cavity fields with randomly distributed scatterers \cite{cavity}, and an integrated array of interferometers \cite{Crespi13}.
Interesting deviations in AL arise when interactions between the particles become significant.
In fact, Anderson himself first noticed the importance of interaction in localization phenomena~\cite{Anderson78} and launched a theoretical investigation in collaboration with Fleishman~\cite{Fleishman80}. 
Recently, advances in technology have reinvigorated theoretical~\cite{theory1,theory2} and experimental~\cite{Aspect09,Sanchez10,Deissler10,Lucioni11,Lahini08, Lahini09,Stutzer12,Naether13} interest on this subject.
For the special case of two interacting particles in a random one-dimensional (1D) potential, Shepelyansky has investigated the interplay of disorder and interaction and concluded that interaction modifies (weakens) localization~\cite{Shepelyansky94}. 
However, several studies analyzing the two-particle case further~\cite{Imry95,Borgonovi95,Oppen96,Romer97,Song,Arias99,Krimer11a,Albrecht12,Dufour12} have shown that the problem of the interplay between disorder and interaction is very complex and the results depend on the details of the system, the localization measure, and the numerical technique employed \cite{Segev13,Sanchez10, Deissler10}. 

We show here that for the two-particle case, a linear two-dimensional (2D) waveguide array is an ideal platform to quantitatively study the role of interaction in AL.
Focusing on bosons, we start by analyzing in detail the general model describing the dynamics of two interacting particles in a disordered 1D lattice.
The effect of on-site interaction on localization is quantified using a measure known as the participation ratio for both diagonal and off-diagonal disorder. We also evaluate the behavior of the particles' second-order correlation function, characterizing the spatial quantum interference between them, and discuss how it can be directly measured through intensity measurements on the output distribution of light from the photonic lattice. We discuss cases showing that the effect of interaction on AL is strongly dependent on the type of disorder and initial conditions, but is independent of whether the interaction is repulsive or attractive. 
The origin of the latter is briefly explained, followed by an explicit example where the indifference is broken. 
We conclude by laying out the details of a photonic implementation in a linear 2D waveguide array, assuming only classical sources of light. Here the on-site interaction can be realized by changing the relative detuning of diagonal waveguides with respect to off-diagonal waveguides as initially proposed in Refs.~\cite{Longhi11a,Krimer11b,Corrielli13}. 
 
The ability to tune the particle interaction, the ease in preparing different initial states in optics, and the advantage of performing the experiment with classical light make our setup ideal for studying the role of interactions in AL of two particles with existing technology.

{\it Two interacting bosons in a disordered 1D lattice.} 
We are interested in the dynamics of two interacting bosons in a disordered 1D lattice as governed by the Bose-Hubbard model
\begin{equation}
\hat{H}=\sum_{j}\epsilon_{j}\hat{a}_{j}^{\dagger} \hat{a}_{j}  -\sum_{\langle j,k\rangle} J_{j,k} (\hat{a}_{j}^{\dagger} \hat{a}_{k}+\hat{a}_{k}^{\dagger} \hat{a}_{j}) +  \frac{U}{2}\sum_{j}\hat{n}_{j}(\hat{n}_{j}-1),
\label{ham}
\end{equation}
where $\hat{a}_{j}^{\dagger} (\hat{a}_{j})$ is the boson creation (annihilation) operator at site $j$, $\epsilon_{j}$ is the on-site potential energy at site $j$, $J_{j,k}$ is the tunneling amplitude between nearest neighbors, and $U$ is the on-site interaction strength.

We concentrate on two types of static disorder~\cite{disordertype}. The first is the diagonal disorder, in which the on-site potential energies $\{ \epsilon_{j} \}$ are uniformly randomized within a finite range $(\bar{\epsilon}- \Delta \epsilon,\bar{\epsilon}+ \Delta \epsilon)$. The second type of disorder is the off-diagonal disorder, in which the tunneling amplitudes $\{ J_{j,k} \}$ are uniformly randomized within a finite range $(\bar{J}- \Delta J,\bar{J}+ \Delta J)$. Here, to stay within an experimentally realizable regime, the range of $\Delta\epsilon$ and $\Delta J$ are restricted such that both $\epsilon_{j}$ and $J_{j,k}$ are always positive. 

Later on, we show how this model can be realized in a 2D photonic lattice and the interplay between disorder and interaction can be observed in such a system. For this purpose, we concentrate on two experimentally measurable quantities. The first is called the participation ratio (PR)~\cite{sd}, defined as ${\rm PR}(t)=(\sum_{j}^{L}\vert\psi_{j}(t)\vert^{2})^{2}/\sum_{j}^{L} \vert \psi_{j}(t) \vert^{4}$, where $\vert\psi_{j}(t)\vert^{2}$ denotes the normalized density of particles at site $j$, given by $\vert\psi_{j}(t)\vert^{2}=\frac{1}{2}\bra{\psi(t)} \hat{n}_{j} \ket{\psi(t)}$, with $\sum_{j}^{L}\vert\psi_{j}(t)\vert^{2}=1$.
It provides a notion of an effective number of occupied sites such that PR$\rightarrow 1$ and PR$\rightarrow L$ for the most localized and completely delocalized states, respectively. 
Here, $\ket{\psi(t)}$ denotes the time-dependent wave function of two bosons, obtained by solving the Schr\"odinger equation.
The second is the intensity correlation function defined as $\Gamma_{j,k}(t)=\langle \hat{a}_{j}^{\dagger} \hat{a}_{k}^{\dagger} \hat{a}_{k} \hat{a}_{j}\rangle$. It characterizes the spatial quantum interference of two particles, known as the Hanbury Brown-Twiss correlation, and can  be used to observe qualitative differences between different types of disorder as in Refs. \cite{Lahini10,Lahini12}. 

{\it Interplay between disorder and interaction.}
First we study the interplay between disorder and interaction using three initial states corresponding to two bosons placed at (i) the same site,  $\frac{1}{\sqrt{2}}(\hat{a}_{0}^{\dagger})^{2}\ket{0}$; (ii) adjacent sites, $\hat{a}_{0}^{\dagger}\hat{a}_{1}^{\dagger}\ket{0}$; and (iii) two sites separated by an empty site, $\hat{a}_{-1}^{\dagger}\hat{a}_{1}^{\dagger}\ket{0}$.

To investigate the AL of two interacting particles, one would generally want to consider a long time scale after which the localization of two particles is saturated. However, in this paper we concentrate on an experimentally realizable time scale \cite{fn}, which is in general shorter than the saturation time. 
The trend of the PR with respect to increasing $U$ (in units of $\bar{J}$) for the time well after saturation was checked to be similar to the final time chosen.

\begin{figure}[b]
\centering
\includegraphics[width=8.8cm]{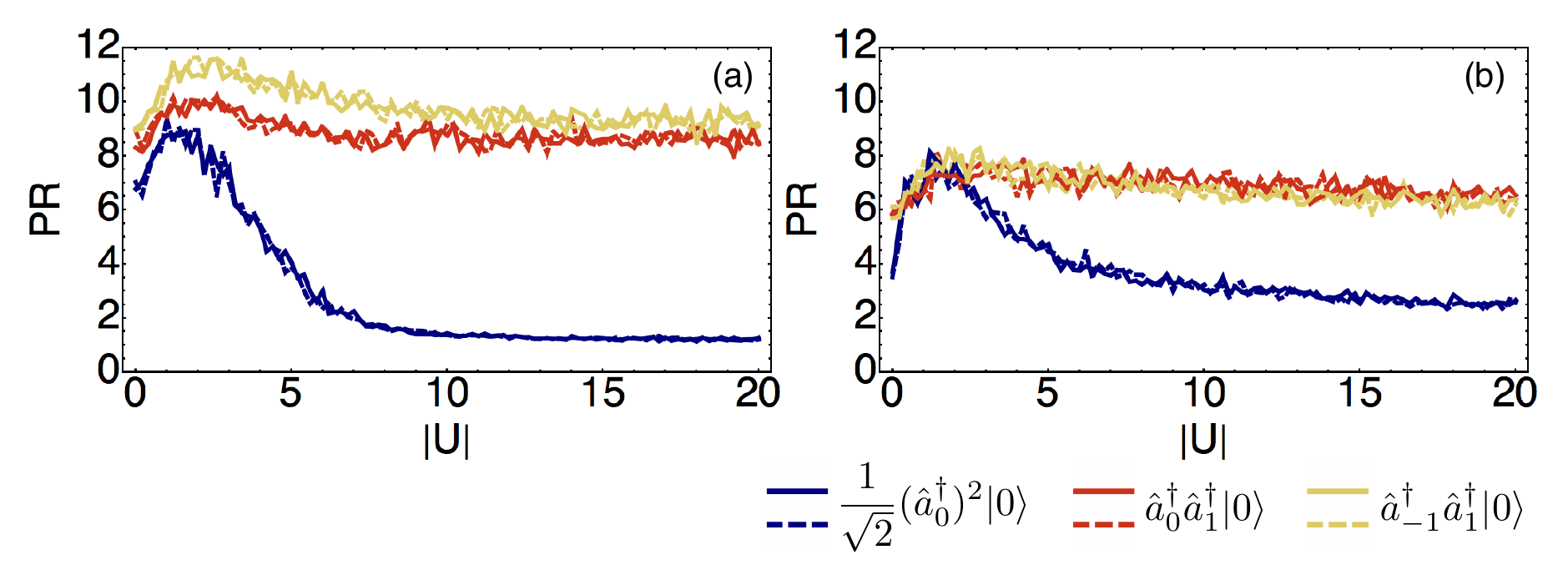}
\caption{ Participation ratio for (a) diagonal disorder: $(\bar{\epsilon}, \Delta\epsilon, \bar{J}, \Delta J)=(2,2,1,0)$, and (b) off-diagonal disorder: $(\bar{\epsilon}, \Delta\epsilon, \bar{J}, \Delta J)=(2,0,1,1)$.
Three initial conditions are considered: $\frac{1}{\sqrt{2}}(\hat{a}_{0}^{\dagger})^{2}\ket{0}$, $\hat{a}_{0}^{\dagger}\hat{a}_{1}^{\dagger}\ket{0}$, and $\hat{a}_{-1}^{\dagger}\hat{a}_{1}^{\dagger}\ket{0}$ correspond to the blue, red, and yellow solid (dashed) curves, respectively, for repulsive (attractive) interaction.}
\label{PR}
\end{figure}

The exact numerical calculations for $\ket{\psi(t)}$ presented in Fig.~\ref{PR} show that the effect of interaction on the AL of two bosons depends on the type of disorder and the initial condition and is independent of whether the interaction is repulsive or attractive for the initial conditions used. For case (i), the interaction suppresses localization within a moderate value of $\vert U \vert$, on the order of disorder strength, for both types of disorder, so that they are less localized compared to the noninteracting case as previously noted~\cite{Shepelyansky94}. 
Increasing $\vert U \vert$ further tends to enhance localization for both types of disorder and especially for the diagonal disorder; two bosons are completely localized (PR$\rightarrow 1$) at the initial position for a large value of $\vert U \vert$. Similar trends have also been reported in a waveguide setup~\cite{Stutzer12, Naether13}, ultracold bosons~\cite{Vermersch12}, and correlated electrons~\cite{electrons} under the mean-field approximation. 
For cases (ii) and (iii), the effects of interaction are less pronounced. The suppression of localization is still there for small $\vert U \vert$, but the large $\vert U \vert$ behavior is quite different from that arising from the same-site initial condition. The effect of the on-site interaction on AL seems to be larger for diagonal disorder than off-diagonal disorder for all the initial conditions. 
Note that such behavior is the result of an interplay between disorder and interaction, each of which cannot yield the same behavior by itself.

\begin{figure}[b]
\centering
\includegraphics[width=7.5cm]{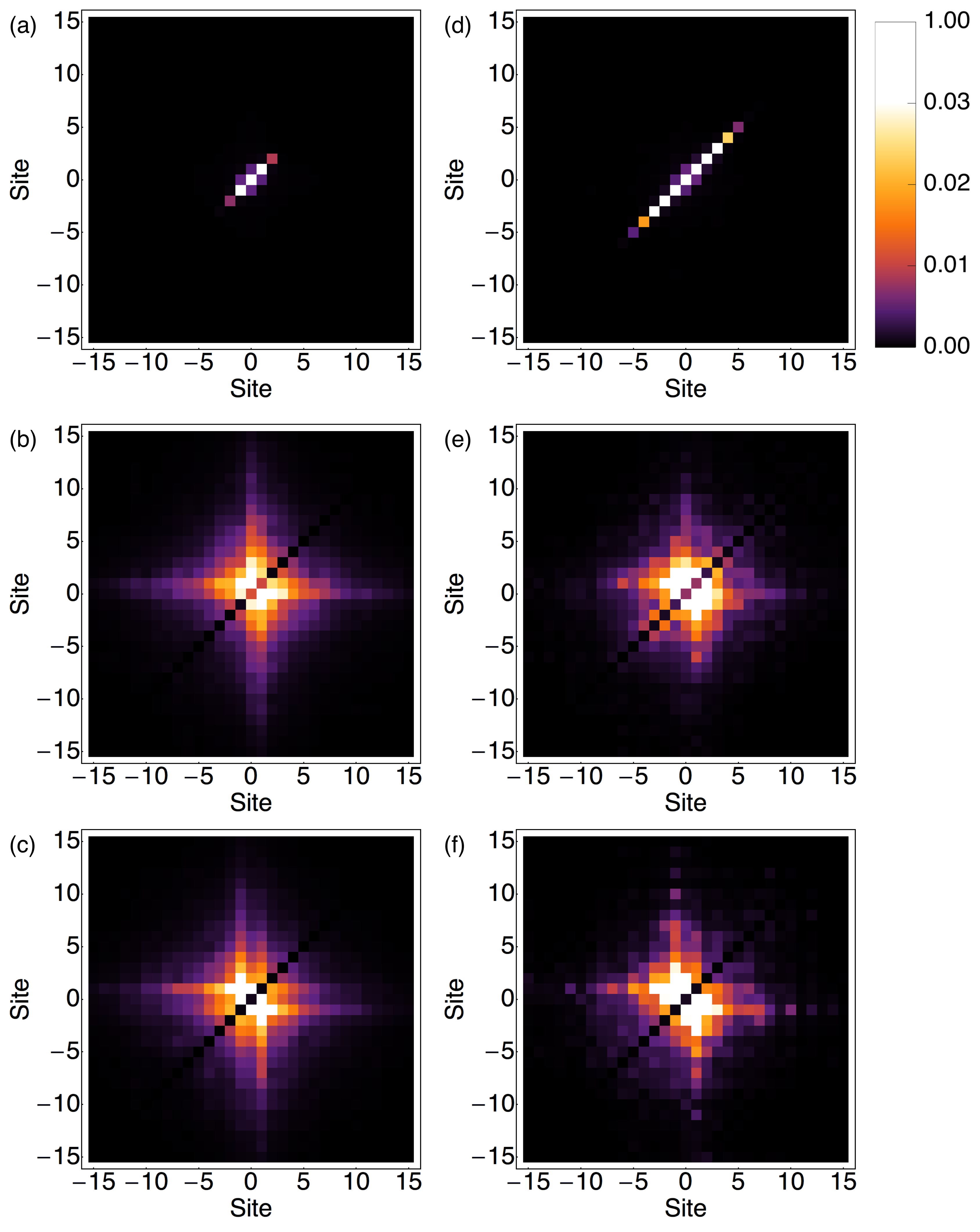}
\caption{Intensity correlation function for diagonal (left column) and off-diagonal (right column) disorder. The initial states are $\frac{1}{\sqrt{2}}(\hat{a}_{0}^{\dagger})^{2}\ket{0}$, $\hat{a}_{0}^{\dagger}\hat{a}_{1}^{\dagger}\ket{0}$, and $\hat{a}_{-1}^{\dagger}\hat{a}_{1}^{\dagger}\ket{0}$ from top to bottom, respectively. Parameters values are as in Fig. 1.
}
\label{correlation}
\end{figure}

In the large $\vert U \vert$ limit, the two non-same-site initial conditions yield almost the same values of PR. However, the differences between conditions (ii) and (iii) are manifest in quantum correlations as shown in Fig.~\ref{correlation}, depicting the correlation functions for $U=20$. The two bosons placed initially at the same site [case (i)] propagate together [see Figs.~\ref{correlation} (a) and (d)], and behave as a bound pair even for repulsive interaction~\cite{Winkler06}. On the other hand, a qualitative difference between the other two initial conditions is found; two bosons can stay together only around their initial position for case (ii) [see Figs.~\ref{correlation} (b) and (e)], but two bosons never stay together for case (iii) [see Figs.~\ref{correlation} (c) and (f)]. Such a difference in quantum correlation is induced by strong interaction and vanishes with decreasing $\vert U\vert$.


{\it Repulsive versus attractive interaction.}
The results shown so far have been independent of whether the interaction is repulsive or attractive. This is a result of the symmetries of the observables and the initial states~\cite{symmetry}. We briefly describe these symmetries and provide an example where the nature of the interaction matters.

The Hamiltonian, Eq.~(\ref{ham}), has the following symmetry. Defining the Hamiltonian $H_{+}$ ($H_{-}$) as the repulsive (attractive) case, the expectation values of an observable under these Hamiltonians $\langle \hat{O}(t) \rangle_\pm$ are identical if both the initial state $\ket{\psi(0)}$ and the observable $\hat{O}$ are invariant under the time reversal and $\pi$-boost transformation~\cite{footnote3}. This is satisfied by the initial states and the observables used in Figs.~\ref{PR} and \ref{correlation}. Thus, to observe the difference between repulsive and attractive interactions, either the initial state or the observable that is not invariant under the time reversal or $\pi$-boost operation is required. 

\begin{figure}[b]
\centering
\includegraphics[width=8.8cm]{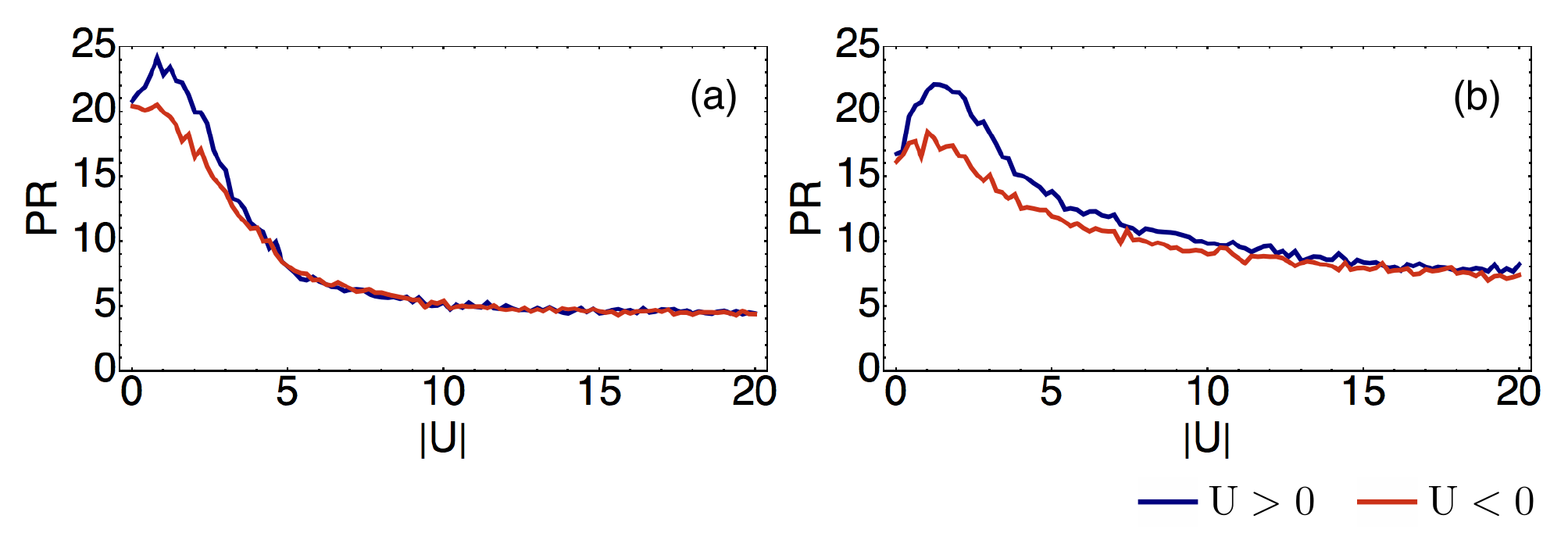}
\caption{ Participation ratio as a function of $U$ for an initial state that breaks the $\pi$-boost symmetry, $\ket{\psi(0)}=\frac{1}{\sqrt{2}} (\frac{1}{\sqrt{2}}(\hat{a}_{0}^{\dagger})^{2} +\hat{a}_{1}^{\dagger}\hat{a}_{2}^{\dagger} )\ket{0}$ (see text):
(a) diagonal disorder, $(\bar{\epsilon}, \Delta\epsilon, \bar{J}, \Delta J)=(2,1,1,0)$; (b) off-diagonal disorder, $(\bar{\epsilon}, \Delta\epsilon, \bar{J}, \Delta J)=(2,0,1,0.5)$. The blue (red) solid curve corresponds to the repulsive (attractive) interaction. The rest of the parameters are as in Fig. 1.
}
\label{entPR}
\end{figure}

As an example, we consider an initial condition that breaks the $\pi$-boost symmetry, $\ket{\psi(0)}=\frac{1}{\sqrt{2}} (\frac{1}{\sqrt{2}}(\hat{a}_{0}^{\dagger})^{2} +\hat{a}_{1}^{\dagger}\hat{a}_{2}^{\dagger} )\ket{0}$.
Figure \ref{entPR} presents the effect of interaction for this initial condition in the presence of weak disorder. Strong disorder was checked to destroy the differences between $\pm U$ for this example. 
In the figure, one can see that the repulsive interaction yields larger PR and that the difference between $\pm U$ diminishes as the interaction strength increases. The latter is expected because $\ket{\psi(0)}$ is time reversal invariant. 

\begin{figure}[hb]
\centering
\includegraphics[width=8.8cm]{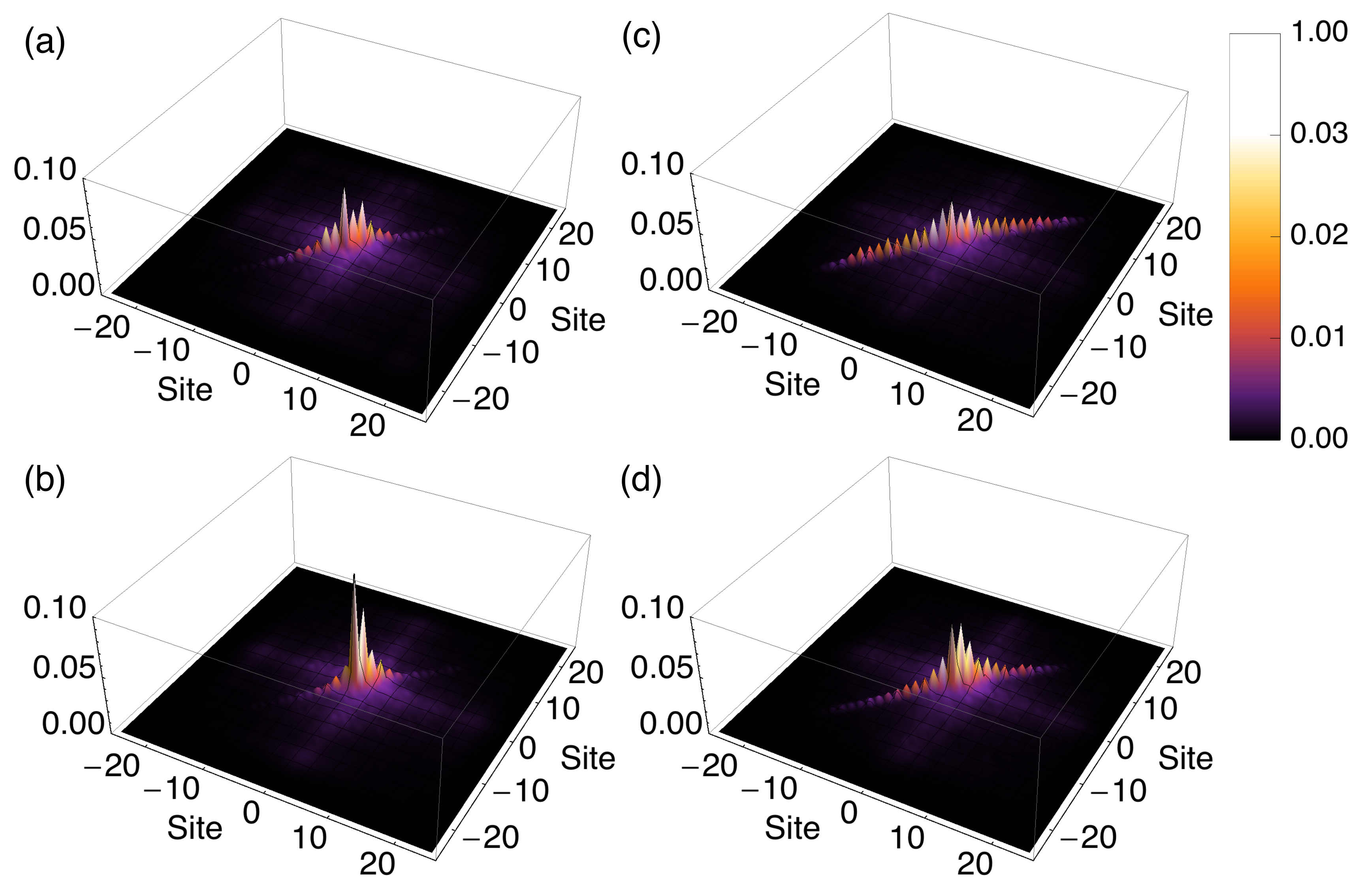}
\caption{Intensity correlation function for diagonal disorder (left column) and off-diagonal disorder (right column), where $U=2$ (top row) and $U=-2$ (bottom row). The rest of the parameters are as in Fig. 3.
}
\label{entcorrelation}
\end{figure}

The correlation functions for $U=\pm 2$ plotted in Fig.~\ref{entcorrelation} show the differences graphically. As expected, the repulsive case [Figs.~\ref{entcorrelation} (a) and (c)] has a broader distribution with a suppressed peak in the middle compared to the attractive case [Figs.~\ref{entcorrelation} (b) and (d)].

{\it  Anderson localization of two interacting particles in linear photonic lattices.}\label{realization}
\begin{figure}[!t]
\centering
\includegraphics[width=8.5cm]{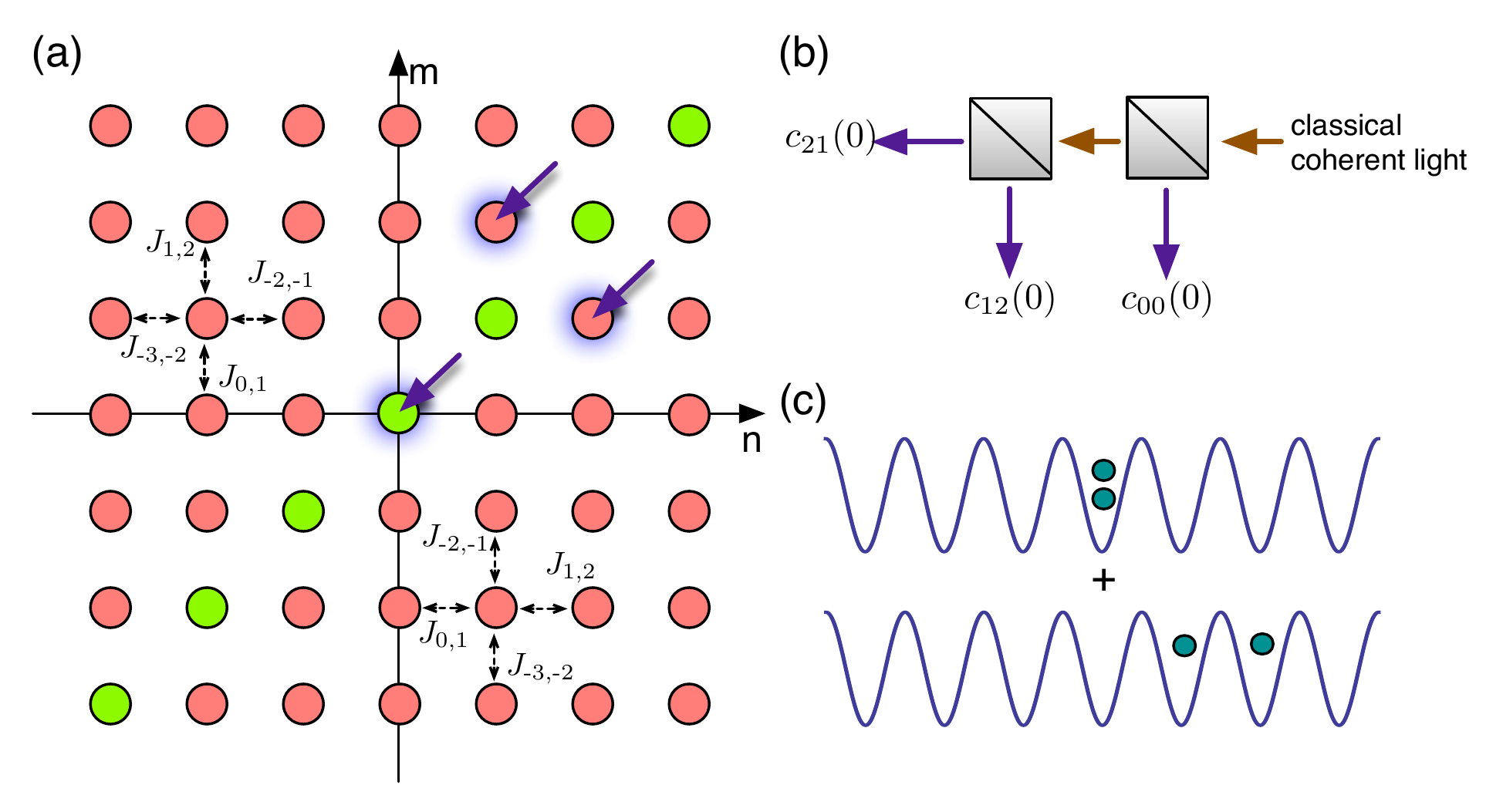}
\caption{
(a) A 2D waveguide array, where, as an example, classical input beams generated by scheme (b) are launched into the waveguides indicated by the purple arrows, corresponding to (c) the initial condition in the 1D lattice $\ket{\psi(0)}=\frac{1}{\sqrt{2}}(\frac{1}{\sqrt{2}}(\hat{a}_{0}^{\dagger})^{2}+\hat{a}_{1}^{\dagger}\hat{a}_{2}^{\dagger})\ket{0}$. The waveguide is symmetric upon reflection along the diagonal.
}
\label{2D}
\end{figure}
Perhaps surprisingly, linear propagation of light in 2D waveguide arrays can be used to experimentally observe two interacting particles in a 1D lattice~\cite{Longhi11a,Krimer11b,Corrielli13}. Now, we briefly explain the connection between the two, concentrating on the bosonic case, and then discuss how such a setup allows experimental observation of our theoretical studies with only classical sources of light. It will be shown that the 2D waveguide array is an ideal platform in many ways to observe the role of interaction in two-particle AL. 

To see the formal similarity between the linear 2D waveguide array and the two interacting particles in 1D, it is helpful to expand the state vector $|\psi (t)\rangle$ of two particles in 1D in Fock space as
\begin{align}
\label{2Dstate}
\ket{\psi(t)} &= \frac{1}{\sqrt{2}}\sum_{n,m=1}^{L} c_{n,m}(t)\hat{a}_{n}^\dagger \hat{a}_{m}^\dagger \ket{0} \nonumber \\
&= \sum_{n=1}^{L} c_{n,n}(t) \frac{1}{\sqrt{2}}(\hat{a}_{n}^\dagger)^{2} \ket{0} \nonumber \\
&~~~~~~~~+\sum_{m>n} \frac{1}{\sqrt{2}}[c_{n,m}(t)+c_{m,n}(t)]\hat{a}_{n}^\dagger \hat{a}_{m}^\dagger \ket{0},
\end{align}
where $c_{n,m}(t)$ is the probability amplitude to find one boson at site $n$ and the other at site $m$, $c_{n,m}(t)=c_{m,n}(t)$ at any $t>0$ due to the indistinguishability of the particles. 
Substituting it into the Schr\"odinger equation, one obtains
\begin{eqnarray}\label{twod}
\begin{split}
i \dot{c}_{n,m}
 &=(\epsilon_{n}+\epsilon_{m}+U \delta_{n,m}) {c}_{n,m}\\
 &\hspace{1cm} -J_{n-1,n}c_{n-1,m}- J_{n,n+1} c_{n+1,m}\\
 &\hspace{2cm} -J_{m-1,m}c_{n,m-1}- J_{m,m+1} c_{n,m+1}
\end{split}
\end{eqnarray}
The dynamics for $c_{n,m}(t)$ described by Eq.~(\ref{twod}) is equivalent to the coupled-mode equations for light propagation in a symmetric square 2D waveguide array, where $c_{n,m}(t)$ describe the amplitudes of the classical field at $(n,m)$ and the reflection symmetry along the diagonal axis holds. This establishes the connection between the model Hamiltonian and the photonic lattice. In the 2D structure, the first term in Eq.~(\ref{twod}), $(\epsilon_{n}+\epsilon_{m}+U \delta_{n,m})$, is determined by the width and the refractive index of the $(n,m)$th waveguide, and the tunneling amplitude $J_{i,j}$ defines the coupling constant between the $i$th and $j$th neighboring waveguides in the $x$ or $y$ axis [see Fig.~\ref{2D} (a), corresponding to the case of seven sites in 1D]. Here, the effective on-site interaction $U$ is achieved by fabricating the diagonal waveguides with a different refractive index or width with respect to off-diagonal waveguides. Note that such a 2D waveguide structure can be experimentally realized by femtosecond laser writing in fused silica, where the defect line is introduced by varying the writing speed of the laser beam~\cite{laser, Grafe12}. Also, the sign of $U$ can be varied by changing the relative detuning between the waveguides \cite{Corrielli13}. In this way, one can obtain a strong interaction regime in the experimentally accessible 2D waveguide geometry without nonlinear materials. 

Randomized on-site energy for the diagonal disorder can be introduced by a controlled variation of the width~\cite{Lahini08} or the refractive index~\cite{Abouraddy12, randompropagation} of each waveguide while keeping the tunneling rates between waveguides constant. Randomized tunneling strength can be introduced by a controlled variation of the separation between adjacent waveguides~\cite{Naether13, Schwartz07, randomseparation}, while keeping the symmetry with respect to the diagonal axis. For this off-diagonal disorder, identical waveguides should be employed except in the diagonal axis where a difference between the refractive indices of the diagonal ($n_{d}$) and off-diagonal ($n_{0}$) waveguides effectively leads to the interaction $U\sim (n_{0}-n_{d})$. Recently, the values of the diagonal disorder $\Delta \epsilon/\bar{\epsilon}$, the off-diagonal disorder $\Delta J/\bar{J}$, and the interaction $U/\bar{J}$ have been shown to be implementable to about $3$~\cite{Lahini08}, $0.91$~\cite{randomseparation}, and $20$~\cite{Longhi11b}, respectively.

The probability amplitude of finding the two particles at site $n$ is given by $p_{n,n}(t)=c_{n,n}(t)$, whereas the probability amplitude of finding them at lattice sites $n$ and $m$ is given by $p_{n,m}(t)=\frac{1}{\sqrt{2}}(c_{n,m}(t)+c_{m,n}(t))=\sqrt{2}c_{n,m}(t)$. Subsequently, the normalized particle-density distribution $P_{n}(t)$ (=$\vert \psi_{n}(t) \vert^{2}$) is defined by $P_{n}(t)=\frac{1}{2}\bra{\psi(t)}\hat{n}_{n}\ket{\psi(t)}=\sum_{m}\vert c_{n,m}(t)\vert^{2}$, with $\sum_{n}P_{n}(t)=1$. Note, therefore, that both the PR and the correlation function~\cite{footnote2} can be simply observed by directly measuring the intensity $\vert c_{n,m}(t) \vert^{2}$ at the $n$th row and $m$th column of the waveguide; i.e., there is no need for coincidence detection as in direct 1D implementations \cite{Keil11}. 
Such an implementation enables one to completely observe the quantum interference of two particles in terms of the correlation function, which could not be observed in a classical nonlinear implementation, where the measured classical correlations for nonlinear waves do not follow the predicted quantum correlations at higher values of $U/\bar{J}$~\cite{Lahini12}.

The initial conditions, $\hat{a}_{n}^{\dagger}\hat{a}_{m}^{\dagger}\ket{0}$, can be prepared with two classical coherent input beams at $(n,m)$ and $(m,n)$ when $n \neq m$ or a single beam at $(n,n)$. 
The initial relative phase between the beams can be chosen to exhibit either the bosonic ($c_{n,m}(0)=c_{m,n}(0)$) or the fermionic ($c_{n,m}(0)=-c_{m,n}(0)$) statistics.
It is then straightforward to realize an arbitrary initial condition for two bosons in a 1D lattice; for example, the initial condition $\ket{\psi(0)}=\frac{1}{\sqrt{2}}(\frac{1}{\sqrt{2}}(\hat{a}_{0}^{\dagger})^{2}+\hat{a}_{1}^{\dagger}\hat{a}_{2}^{\dagger})\ket{0})$ can be realized by choosing $c_{0,0}(0)=1/\sqrt{2}$ and $c_{1,2}(0)=c_{2,1}(0)=1/2$ (see Fig.~\ref{2D}). Such an initial state can be simply generated by an appropriate combination of $50/50$ beam splitters~\cite{bs} as shown in Fig.~\ref{2D}(b). More generally, an arbitrary initial condition can be generated by a properly controlled multiport beam splitter into which a classical coherent light is initially launched~\cite{mb}. 

With the implementation scheme described above, the AL of two interacting bosons can be experimentally studied in the linear 2D waveguide array and observables such as the PR and correlation functions can be straightforwardly measured. In particular, an arbitrary initial condition for two bosons can be realized with classical coherent input beams.


{\it Conclusions.}
We have shown that a linear 2D photonic lattice using classical sources of light provides an ideal playground for probing the effect of on-site interaction on the AL of two bosons in a 1D disordered lattice.
We have carried out the numerical calculations for two types of disorders, diagonal and off-diagonal, and different initial states under different interaction strengths and signs. 
The effect was characterized in terms of the participation ratio and the correlation function, which clearly display, quantitatively and qualitatively, the interplay between interaction and disorder, and can be measured by an intensity output distribution of 2D arrays.  

{\it Acknowledgments.} C. Lee thanks K. Yee for discussions. 
This work was supported by the National Research Foundation and Ministry of Education (partly through the Tier 3 Grant "Random Numbers from Quantum Processes"), Singapore. We also acknowledge partial travel support by the EU IP-SIQS.


\begin{thebibliography}{99}

\bibitem{Anderson58} P. Anderson, Phys. Rev. {\bf 109}, 1492 (1958).

\bibitem{Schwartz07} T. Schwartz, G. Bartal, S. Fishman, and M. Segev, Nature (London) {\bf 446}, 52 (2007); 
\bibitem{Segev13} M. Segev, Y. Silberberg, and D. N. Christodoulides, Nat. Photonics {\bf 7}, 197 (2013).

\bibitem{Billy08} J. Billy, V. Josse, Z. Zuo, A. Bernard, B. Hambrecht, P. Lugan, D. Cl\'ement, L. Sanchez-Palencia, P. Bouyer, and A. Aspect, Nature (London) {\bf 453}, 891 (2008).
\bibitem{Roati08} G. Roati, C. D'Errico, L. Fallani, M. Fattori, C. Fort, M. Zaccanti, G. Modugno, M. Modugno, and M. Inguscio, Nature (London) {\bf 453}, 895 (2008).
\bibitem{cavity}
R. Dalichaouch, J. P. Armstrong, S. Schultz, P. M. Platzman, and S. L. Mccall, Nature (London) {\bf 354}, 53 (1991); C. Dembowski, H.-D. Gr\"af, R. Hofferbert, H. Rehfeld, A. Richter, and T. Weiland, Phys. Rev. E {\bf 60}, 3942 (1999); J. D. Bodyfelt, M. C. Zheng, T. Kottos, U. Kuhl, and H.-J. St\"ockmann, Phys. Rev. Lett. {\bf 102}, 253901 (2009).

\bibitem{Crespi13} A. Crespi, R. Osellame, R. Ramponi, V. Giovannetti, R. Fazio, L. Sansoni, F. D. Nicola, F. Sciarrino, and P. Mataloni, Nat. Photonics {\bf 7}, 322 (2013).

\bibitem{Anderson78} P. W. Anderson, Science {\bf 201}, 307 (1978).
\bibitem{Fleishman80} L. Fleishman and P. W. Anderson, Phys. Rev. B {\bf 21}, 2366 (1980).

\bibitem{theory1} A. S. Pikovsky and D. L. Shepelyansky, Phys. Rev. Lett. {\bf 100}, 094101 (2008); S. Flach, D. O. Krimer, and Ch. Skokos, {\it ibid}. {\bf 102}, 024101 (2009); Ch. Skokos, I. Gkolias, and S. Flach, {\it ibid}. {\bf 111}, 064101 (2013).

\bibitem{theory2} C. D'Errico, M. Moratti, E. Lucioni, L. Tanzi, B. Deissler, M. Inguscio, G. Modugno, M. B. Plenio, and F. Caruso, New J. Phys. {\bf 15}, 045007 (2013).


\bibitem{Aspect09} A. Aspect and M. Inguscio, Phys. Today {\bf 62}, 30 (2009).

\bibitem{Sanchez10} L. Sancheze-Palencia and M. Lewenstein, Nat. Phys. {\bf 6}, 87 (2010).

\bibitem{Deissler10} B. Deissler, M. Zaccanti, G. Roati, C. D'Errico, M. Fattori, M. Modugno, G. Modugno, and M. Inguscio, Nat. Phys. {\bf 6}, 354 (2010).

\bibitem{Lucioni11} E. Lucioni, B. Deissler, L. Tanzi, G. Roati, M. Zaccanti, M. Modugno, M. Larcher, F. Dalfovo, M. Inguscio, and G. Modugno, Phys. Rev. Lett. {\bf 106}, 230403 (2011)

\bibitem{Lahini08} Y. Lahini, A. Avidan, F. Pozzi, M. Sorel, R. Morandotti, D. N. Christodoulides, and Y. Silberberg, Phys. Rev. Lett. {\bf 100}, 013906 (2008).
\bibitem{Lahini09} Y. Lahini, R. Pugatch, F. Pozzi, M. Sorel, R. Morandotti, N. Davidson, and Y. Silberberg, Phys. Rev. Lett. {\bf 103}, 013901 (2009).
\bibitem{Stutzer12} S. St\"utzer, Y. V. Kartashov, V. A. Vysloukh, A. T\"unnermann, S. Nolte, M. Lewenstein, L. Torner, and A. Szameit, Opt. Lett. {\bf 37}, 1715 (2012).
\bibitem{Naether13} U. Naether, S. Rojas-Rojas, A. J. Mart\'inez, S. St\"utzer, A. T\"unnermann, S. Nolte, M. I. Molina, R. A. Vicencio, and A. Szameit, Opt. Exp. {\bf 21}, 927 (2013).

\bibitem{Shepelyansky94} D. L. Shepelyansky, Phys. Rev. Lett. {\bf 73}, 2607 (1994).

\bibitem{Imry95} Y. Imry, Europhys. Lett. {\bf 30}, 405 (1995).

\bibitem{Borgonovi95} F. Borgonovi and D. L. Shepelyansky, Nonlinearity {\bf 8}, 877 (1995).
\bibitem{Oppen96} F. von Oppen, T. Wettig, and J. M\"uller, Phys. Rev. Lett. {\bf 76}, 491 (1996).
\bibitem{Romer97} R. A. R\"omer and M. Schreiber, Phys. Rev. Lett. {\bf 78}, 515 (1997); 
K. Frahm, A. M\"uller-Groeling, J.-L. Pichard, and D. Weinmann, Phys. Rev. Lett. {\bf 78}, 4889 (1997);
R. A. R\"omer and M. Schreiber, Phys. Rev. Lett. {\bf 78}, 4890 (1997).
\bibitem{Song} P. H. Song and D. Kim, Phys. Rev. B {\bf 56} 12217 (1997); P. H. Song and F. von Oppen, {\it ibid}. {\bf 59}, 46 (1999).
\bibitem{Arias99} S. De T. Arias, X. Waintal, and J.-L. Pichard, Eur. Phys. J. B {\bf 10}, 149 (1999).
\bibitem{Krimer11a} D. O. Krimer, R. Khomeriki, and S. Flach, JETP Letters {\bf 94}, 406 (2011).
\bibitem{Albrecht12} C. Albrecht and S. Wimberger, Phys. Rev. B {\bf 85}, 045107 (2012).
\bibitem{Dufour12} G. Dufour and Giuliano Orso, Phys. Rev. Lett. {\bf 109}, 155306 (2012).

\bibitem{Longhi11a} S. Longhi, Opt. Lett. {\bf 36}, 3248 (2011).

\bibitem{Krimer11b} D. O. Krimer and R. Khomeriki, Phys. Rev. A {\bf 84}, 041807(R) (2011).

\bibitem{Corrielli13} G. Corrielli, A. Crespi, G. D. Valle, S. Longhi, and R. Osellame, Nat. Commun. {\bf 4}, 1555 (2013).

\bibitem{disordertype} Y. Lahini, Y. Bromberg, Y. Shechtman, A. Szameit, D. N. Christodoulides, R. Morandotti, and Y. Silberberg, Phys. Rev. A {\bf 84}, 041806(R) (2011); J. B. Pendry, J. Phys. C: Solid State Phys. {\bf 15}, 5773 (1982).

\bibitem{sd} We do not use the second moment, which is one of the most commonly used measures for the description of localization phenomena and appropriate to the distribution localized at the center.  In this work, it may not be localized at the center depending on the initial condition even when it is completely localized.

\bibitem{Lahini10} Y. Lahini, Y. Bromberg, D. N. Christodoulides, and Y. Silberberg, Phys. Rev. Lett. {\bf 105}, 163905 (2010).

\bibitem{Lahini12} Y. Lahini, M. Verbin, S. D. Huber, Y. Bromberg, R. Pugatch, and Y. Silberberg, Phys. Rev. A {\bf 86}, 011603(R) (2012).

\bibitem{fn} We use $t=10$ (all units are defined with respect to $\bar{J}$ throughout the paper) for the final time and 51 lattices throughout the paper. All the results are averages over 100 realizations, obtained by solving the Schr\"odinger equation with exact diagonalization. Edge effects were checked to be negligible for all the figures. The results for the short time scale ($t=10$) were checked to be qualitatively similar to those of the extremely long time scale ($t=10^{5}$).

\bibitem{Vermersch12} B. Vermersch and J. C. Garreau, Phys. Rev. E {\bf 85}, 046213 (2012).

\bibitem{electrons} T. Vojta, F. Epperlein, and M. Schreiber, Phys. Rev. Lett. {\bf 81}, 4212 (1998); K. Byczuk, W. Hofstetter, U. Yu, and D. Vollhardt, Eur. Phys. J.: Spec. Top. {\bf 180}, 135 (2010).

\bibitem{Winkler06} K. Winkler, G. Thalhammer, F. Lang, R. Grimm, J. Hecker Denschlag, A. J. Daley, A. Kantian, H. P. B\"uchler, and P. Zoller, Nature (London) {\bf 441}, 853 (2006).

\bibitem{symmetry} K. Sakmann, A. I. Streltsov, O. E. Alon, and L. S. Cederbaum, Phys. Rev. A {\bf 82}, 013620 (2010); U. Schneider, L. Hackerm\"uller, J. P. Ronzheimer, S. Will, S. Braun, T. Best, I. Bloch, E. Demler, S. Mandt, D. Rasch, and A. Rosch, Nat. Phys. {\bf 8}, 213 (2012).

\bibitem{footnote3}
The time reversal operator $\hat{R}_{t}$ that transforms the wave function into its complex conjugate equivalently modifies the time evolution operator : $\hat{R}_{t}e^{-i\hat{H}t}\hat{R}_{t}^{\dagger}=e^{i\hat{H}t}$.
The $\pi$-boost operator $\hat{B}$ assigns an additional position-dependent phase in real space : $\hat{B}\hat{a}_{j}\hat{B}=e^{i\pi j}\hat{a}_{j}$.

\bibitem{laser} D. Bl\"omer, A. Szameit, F. Dreisow, T. Schreiber, S. Nolte, and A. T\"unnermann, Opt. Express {\bf 14}, 2151 (2006); 
A. Szameit, F. Dreisow, T. Pertsch, S. Nolte, and A. T�nnermann, Opt. Express {\bf 15}, 1579 (2007);  
A. Szameit and S. Nolte, J. Phys. B: At. Mol. Opt. Phys. {\bf 43}, 163001 (2010).

\bibitem{Grafe12} M. Gr\"afe, A. S. Solntsev, R. Keil, A. A. Sukhorukov, M. Heinrich, A. T\"unnermann, S. Nolte, A. Szameit, and Yu S. Kivshar, Sci. Rep. {\bf 2}, 562 (2012).

\bibitem{randompropagation} A. Szameit, Y. V. Kartashov, P. Zeil, F. Dreisow, M. Heinrich, R. Keil, S. Nolte, A. T\"unnermann, V. A. Vysloukh, and L. Torner, Opt. Lett. {\bf 35}, 1172 (2010).

\bibitem{Abouraddy12} A. F. Abouraddy, G. Di Giuseppe, D. N. Christodoulides, and B. E. A. Saleh, Phys. Rev. A {\bf 86}, 040302(R) (2012).

\bibitem{randomseparation} L. Martin, G. Di Giuseppe, A. Perez-Leija, R. Keil, Felix Dreisow, M. Heinrich, S. Nolte, A. Szameit, A. F. Abouraddy, D. N. Christodoulides, and B. E. A. Saleh, Opt. Exp. {\bf 19}, 13636 (2011); M. Heinrich, R. Keil, Y. Lahini, U. Naether, F. Dreisow, A. T\"unnermann, S. Nolte and A. Szameit, New J. Phys. {\bf 14}, 073026 (2012).

\bibitem{Longhi11b} S. Longhi and G. D. Valle, Opt. Lett. {\bf 36}, 4743 (2011).

\bibitem{footnote2} 
The correlation functions, $\Gamma_{n,m}=\langle \hat{a}_{n}^{\dagger} \hat{a}_{m}^{\dagger} \hat{a}_{m} \hat{a}_{n}\rangle$, can be rewritten for two-particle cases in terms of the probability of finding the two particles $\Gamma_{n,m}=2 \vert p_{n,n} \vert^{2}$ for $n=m$ and $\Gamma_{n,m}=\vert p_{n,m}\vert^{2}$ for $n\neq m$.

\bibitem{Keil11} R. Keil, A. Perez-Leija, F. Dreisow, M. Heinrich, H. Moya-Cessa, S. Nolte, D. N. Christodoulides, and A. Szameit, Phys. Rev. Lett. {\bf 107}, 103601 (2011); R. Keil, F. Dreisow, M. Heinrich, A. T\"unnermann, S. Nolte, and A. Szameit, Phys. Rev. A {\bf 83}, 013808 (2011).

\bibitem{bs} Here, the beam-splitter operation is represented by $\hat{a}^{\dagger} \rightarrow (\hat{c}^{\dagger}+\hat{d}^{\dagger})/\sqrt{2}$ and $\hat{b}^{\dagger} \rightarrow (\hat{c}^{\dagger}-\hat{d}^{\dagger})/\sqrt{2}$, where $\hat{a}^{\dagger}$ ($\hat{c}^{\dagger}$) and �$\hat{b}^{\dagger}$ ($\hat{d}^{\dagger}$) denote the input (output) modes.

\bibitem{mb} M. Reck, A. Zeilinger, H. J. Bernstein, and P. Bertani, Phys. Rev. Lett. {\bf 73}, 58 (1994); M. \.Zukowski, A. Zeilinger, and M. A. Horne, Phys. Rev. A {\bf 55}, 2564 (1997); K. Mattle, M. Michler, H. Weinfurter, A. Zeilinger, and M. \.Zukowski, Appl. Phys. B {\bf 60}, S111 (1995).



\end{thebibliography}
\end{document}